\documentclass[
  aps, 
  prx, 
  reprint,
  showkeys,
  nobibnotes,
]{revtex4-2}

\usepackage{enumitem}
\setlist[enumerate,1]{label=\textbf{(\roman*)}}
\setlist{
  leftmargin=.65cm,
}

\usepackage{lmodern}
\usepackage{anyfontsize}
    
\usepackage{multirow}
\usepackage{hhline}

\usepackage{tabularray}
\UseTblrLibrary{booktabs}

\usepackage{adjustbox}
\usepackage{tcolorbox}

\newtcolorbox{standout}{
  colback=gray!15,
  boxrule=0pt,
  left=.3cm,
  right=.3cm,
  top=.18cm,
  bottom=.18cm,
  boxsep=0pt
}

\usepackage{mathtools}
\usepackage{amssymb}
\usepackage{amsthm}
\usepackage{xfrac}
\usepackage{stmaryrd} 
\usepackage{scalerel} 
\usepackage{stackengine} 

 \newcommand{\bracket}[3]{%
  \stretchleftright
    {#1}
    {%
      \ensurestackMath{\addstackgap[1pt]{#2}}%
      \vrule width 0pt depth 2pt height 0pt
    }
    {#3}%
} 
\newcommand{\scaledbracket}[3]{%
  \ThisStyle{%
    \stretchleftright
      {#1}
      {
        \ensurestackMath{\addstackgap[1pt]{\SavedStyle #2}}%
        \vrule width 0pt depth 1.5pt height 0pt
      }
      {#3}%
  }%
}

\theoremstyle{plain}
\newtheorem{theorem}{Theorem}[section]

\newtheorem{proposition}[theorem]{Proposition}

\theoremstyle{definition}

\theoremstyle{remark}

\usepackage[
  colorlinks=true,
  linkcolor=darkgreen,
  citecolor=darkgreen,
  urlcolor=darkgreen
]{hyperref}
\usepackage{xurl} 

\usepackage{cleveref}
\crefname{equation}{}{}
\crefname{section}{\S}{\S\S}
\crefname{subsection}{\S}{\S\S}
\crefname{subsubsection}{\S}{\S\S}
\crefname{definition}{Def.}{Defs.}
\crefname{theorem}{Thm.}{Thms.}
\crefname{corollary}{Cor.}{Cors.}
\crefname{lemma}{Lem.}{Lems.}
\crefname{proposition}{Prop.}{Props.}
\crefname{remark}{Rem.}{Rems.}
\crefname{notation}{Ntn.}{Ntns.}
\crefname{fact}{Fact}{Fact}
\crefname{example}{Ex.}{Exs.}
\crefname{figure}{Fig.}{Figs.}
\crefname{table}{Tab.}{Tabs.}
\crefname{footnote}{ftn.}{ftns.}
\Crefname{footnote}{Ftn.}{Ftns.}

\usepackage{tikz}
\usetikzlibrary{
  cd,
  calc,
  arrows.meta,
  backgrounds,
  decorations,
  decorations.pathmorphing, 
}

\tikzcdset{
  arrow style=tikz, 
  diagrams={
    trim left=
    {([xshift=5pt]current bounding box.west)},
    trim right=
    {([xshift=-5pt]current bounding box.east)},
    baseline=
    {([yshift=-2pt]current bounding box.center)},
    >={
      Computer Modern Rightarrow[
        length=4pt, width=4pt
      ]
    },
    hook/.style={
      {Hooks[right, length=2pt, width=8pt]}->
    },
    hook'/.style={
      {Hooks[left, length=2pt, width=8pt]}->
    },
    shorten=0pt,
  }
}

\definecolor{darkblue}{rgb}{0.05,0.25,0.65}
\definecolor{darkgreen}{RGB}{20,140,10}
\definecolor{lightgray}{rgb}{0.9,0.9,0.9}
\definecolor{darkorange}{RGB}{200,100,5}
\definecolor{darkyellow}{rgb}{.91,.91,0}
\definecolor{lightolive}{RGB}{225, 220, 185}

\let\originalsslash\sslash
\renewcommand{\sslash}{\mathord{\originalsslash}}

\makeatletter
\newcommand{\cpt}{\mathpalette\cpt@inner\relax}
\newcommand{\cpt@inner}[2]{%
  \scalebox{0.5}[0.9]{$#1\cup$}
  #1\{\infty\}
}
\makeatother

\DeclareRobustCommand{\rchi}{{\mathpalette\irchi\relax}}
\newcommand{\irchi}[2]{\raisebox{\depth}{$#1\chi$}} 

\tikzset{
  snake left/.style={
    rounded corners,
    to path={
      let \p1 = (\tikztostart.east),
          \p2 = (\tikztotarget.west),
          \p3 = ($(\p1)!0.5!(\p2)$),
          \n1 = {8pt} 
      in
      (\p1)
      -- (\x1 + \n1, \y1)
      -- (\x1 + \n1, \y3)
      -- (\x2 - \n1, \y3) \tikztonodes
      -- (\x2 - \n1, \y2)
      -- (\p2)
    }
  }
}

\tikzset{
  uphordown/.style={
    rounded corners,
    to path={
      let \p1 = (\tikztostart.north),
          \p2 = (\tikztotarget.north),
          \n1 = {max(\y1,\y2) + 8pt}
      in
      (\p1)
      -- (\x1, \n1)
      -- (\x2, \n1) \tikztonodes 
      -- (\p2)
    }
  }
}

\tikzset{
  downhorup/.style={
    rounded corners,
    to path={
      let \p1 = (\tikztostart.south),
          \p2 = (\tikztotarget.south),
          \n1 = {min(\y1,\y2) - 8pt}
      in
      (\p1)
      -- (\x1, \n1)
      -- (\x2, \n1) \tikztonodes 
      -- (\p2)
    }
  }
}

\tikzset{
  rightvertleft/.style={
    rounded corners,
    to path={
      let \p1 = (\tikztostart.east),
          \p2 = (\tikztotarget.east),
          \n1 = {max(\x1,\x2) + 8pt}
      in
      (\p1)
      -- (\n1, \y1)
      -- (\n1, \y2) \tikztonodes 
      -- (\p2)
    }
  }
}

\tikzset{
  leftvertright/.style={
    rounded corners,
    to path={
      let \p1 = (\tikztostart.west),
          \p2 = (\tikztotarget.west),
          \n1 = {min(\x1,\x2) - 8pt}
      in
      (\p1)
      -- (\n1, \y1)
      -- (\n1, \y2) \tikztonodes 
      -- (\p2)
    }
  }
}

\newcommand{\inlinetikzcd}[1]{\begin{tikzcd}[sep=small, ampersand replacement=\&]#1\end{tikzcd}}

\renewcommand{\setminus}{-}

\newcommand{\defneq}{\equiv}

\newcommand{\HilbertSpace}{%
  \mathcal{H}%
}

\begin{document}

\setlength{\abovedisplayskip}{4.5pt}
\setlength{\belowdisplayskip}{4.5pt}
\setlength{\abovedisplayshortskip}{-4pt}
\setlength{\belowdisplayshortskip}{2pt}

\title
{
  FQH Liquids with Flux-Expulsion Islands allow Nonabelian Anyons
}


\author{Hisham Sati}
\email{hsati@nyu.edu}

\author{Urs Schreiber}
\email{us13@nyu.edu}

\affiliation{Mathematics Program and Center for Quantum and Topological Systems,\\
New York University Abu Dhabi, UAE}



\date{\today}


\begin{abstract}
The idea that topologically protected quantum states, such as anyons, may be attached to super/semiconductor heterostructures has received enormous attention, but experimental signatures in 1D systems remain elusive. Here we revisit theoretical underpinnings of anyons in 2D fractional quantum Hall (FQH) systems, whose signatures have been experimentally observed by independent groups. 
We highlight novel theorems about the \emph{Hopfion} or \emph{$\mathbb{C}P^1$-model}, understood as quantization of FQH surplus flux in 2-Cohomotopy, which retrodicts abelian FQH anyon properties in fine detail. Examining this theoretical model further, we find that islands of expelled (surplus) flux enlarge the ground-state monodromy from an abelian group to a (framed spherical) braid group, nonabelian for two or more islands, so that nonabelian anyonic states become topologically admissible; superconducting islands are one candidate realization of such flux expulsion.
\end{abstract}

\maketitle

\section{Introduction}

Anyonic topological order (\cite{Wen1995}, cf. \cite[\S II.B]{SS25-FQAH}) could be the holy grail of quantum materials research (cf. \cite{Stanescu2020,Simon2023}): Degenerate gapped quantum ground states unitarily transforming under adiabatic parameter monodromy. Besides their fundamental theoretical interest, such materials are candidate platforms for intrinsically stabilized quantum computing hardware (cf. \cite{Nayak2008}), possibly necessary for mitigating the decoding bottleneck faced by quantum error correction at scale (cf. \cite{Waintal2024}).

Among proposals for where to find anyonic topological order in materials (as opposed to in quantum simulations), only one stands out to date for having been experimentally observed with confidence (\cite{Nakamura2020}, cf. \cite{Veillon2024}): Fractional quantum Hall systems (FQH, cf. \cite{Stormer99,PapicBalram2024}). These are cold electron liquids confined to effectively 2-dimensional surfaces (typically semiconductor junctions) and subject to transverse magnetic fields so strong and yet so fine-tuned that there is an odd integer (generally rational) number $1/\nu$ of flux quanta per electron. 

Against the backdrop of such a strongly correlated electron liquid, heuristically understood \cite{Jain2007} as a sea of emergent \emph{composite fermion} bound states of electrons and flux quanta, every surplus flux quantum appears as the lack of the $\nu$-th fraction of an electron, thus called a fractional \emph{quasi-hole}. The adiabatic movement of these quasi-holes is theoretically predicted \cite{Arovas1984} to transform the ground state wave function $\vert \psi \rangle \in \HilbertSpace$ of the entire system by a complex phase 
\begin{equation}
  \label{TheBraidingPhase}
  \zeta 
    = 
  e^{\pi \mathrm{i}\nu\, n}
  \in 
  \mathrm{U}(1)
  \mathrlap{\,,}
\end{equation}
which depends solely on the topological \emph{crossing number} $n \in \mathbb{Z}$ of their worldlines, robust against local perturbations of the process.

This is the hallmark property of \emph{topological order}, albeit at this point only in the abelian case (with $\zeta \in \mathrm{U}(1) \subset \mathrm{U}(\HilbertSpace)$). Remarkably, this abelian \emph{anyon braiding phase} has been experimentally observed in recent years by independent groups (\cite{Nakamura2020}, cf. \cite{Veillon2024}).

On the other hand, notably for their potential application as \emph{universal quantum gates}, one is interested (cf. \cite{Nayak2008}) in identifying variant processes in FQH systems which may transform the ground states via \emph{nonabelian} groups of unitaries, $G \subset \mathrm{U}(\HilbertSpace)$. 
A famous proposal argued \cite{DasSarma2005} that this could happen for quasi-holes at $\nu = 5/2$, but industry interest waned when quantum decoherence was found \cite{Inoue2014} to be overwhelming in these systems. 

This has led the community to shift focus to topological states that are argued \cite{Lutchyn2010,Oreg2010} to appear in 1-dimensional semi/superconductor heterostructures (``Majorana zero modes''). However, despite immense effort and prominent claims, conclusive experimental verification of this proposal remains elusive (cf. \cite{DasSarma2023,Kouwenhoven2025}).

An alternative that may deserve more attention is the possibility of nonabelian anyonic states induced by junctions between 2-dimensional abelian FQH liquids and superconductors, which has been variously argued for in \cite{Lindner2012,clarke2013,Vaezi2013,Mong2014,Kim2017,Gul2022,Yoshitome2025}.

In order to put this proposal, of nonabelian anyons induced by superconducting islands in 2D abelian FQH liquids, on robust theoretical foundations, here we revive a possibly underappreciated global formulation of the effective (infrared) nature of topological states: via ``Hopfions'' \cite{Wilczek1983} (review in \cite[\S II.C]{Forte1992}) which we have recently developed further \cite{SS25-AbelianAnyons,SS25-WilsonLoops,SS25-FQH} in application to FQH systems (survey in \cite{SS25-ISQS29}) and also to their crystalline FQAH versions \cite{SS25-FQAH,SS25-Crys}.

\section{Methods}
\label{Methods}

Our methods \cite{SS25-AbelianAnyons,SS25-WilsonLoops,SS25-FQH} of effective anyon field theory are decidedly different (in particular: non-Lagrangian \cite{SS25-ISQS29}) from the traditional approach (recalled and critiqued in \cref{CritiqueOfEffectiveLagrangians} below) which is based on (Chern-Simons type) Lagrangian densities: 

Instead of relegating the global topology of the effective fields to an afterthought on an intrinsically local Lagrangian description, we instead make the relevant choice of topological \emph{flux quantization} \cite{SS25-Flux} (here: of the surplus FQH flux quanta inducing vortex quasi-holes in the electron liquid) the starting point of the discussion (in \cref{OnProperFluxQuantization}). From this one may systematically and rigorously compute \cite{SS25-Complete} the topological quantum observables of the system, using mathematical tools of algebraic topology and representation theory.

The key to this approach is that with a classifying space $\mathcal{A}$ for surplus flux chosen we have that  \cite[\S 2]{SS25-FQH}:

(i) the moduli space of topological parameters for a system inhabiting a surface $\Sigma^2$ is the pointed mapping space $\mathrm{Map}^\ast\bracket({\Sigma^2_{\cpt}, \mathcal{A}})$ homotopy quotiented by the diffeomorphism group $\mathrm{Diff}\bracket({\Sigma^2})$, 

(ii) the adiabatic parameter monodromy is hence the fundamental group $\pi_1 \mathrm{Map}^\ast\bracket({\Sigma^2_{\cpt}, \mathcal{A}}) \sslash \mathrm{Diff}\bracket({\Sigma^2})$,

(iii) and hence the adiabatic transformations of the Hilbert space $\HilbertSpace$ of quantum ground states, characterizing the system's topological quantum order, form unitary representations
$
  \begin{tikzcd}[sep=small]
    \pi_1 
    \bracket({
    \mathrm{Map}^\ast\bracket({\Sigma^2_{\cpt}, \mathcal{A}})
    \sslash
    \mathrm{Diff}\bracket({\Sigma^2})
    })
    \ar[r]
    &
    \mathrm{U}(\HilbertSpace)
    \mathrlap{\,.}
  \end{tikzcd}
$

\subsection{Critique of Effective Lagrangians}
\label{CritiqueOfEffectiveLagrangians}

With a careful re-analysis of FQH anyon quantum states in mind, we critically review a couple of commonly stated but, we argue, insufficiently justified steps in the literature:

It is common to model FQH liquids at long wavelengths by an effective Chern-Simons type Lagrangian density of the form \cite[(2.11)]{Wen1995}\cite[(7.3.10)]{Wen2007}:
\begin{equation}
  \label{EffectiveCSLagrangian}
  L
  :=
  \tfrac{k}{2}
  a\wedge \mathrm{d}a
  -
  A \wedge \mathrm{d}a
  -
  a \wedge j
  \mathrlap{\,,}
\end{equation}
where $A$ is the electromagnetic gauge potential with flux density $F := \mathrm{d}A$, while $J := \mathrm{d}a$ models the electron current density,
and $j$ the quasi-particle current density. This \emph{Ansatz} is justified by the fact that the 
resulting Euler-Lagrange equation of motion 
\begin{equation}
  \label{EffectiveEoM}
  \frac
    { \delta L }
    { \delta a }
  =
  0
  \;\;\;
  \Leftrightarrow
  \;\;\;
  J
  =
  \tfrac{1}{k}
  \bracket({
    F + j
  })
\end{equation}
correctly expresses, at filling fraction $\nu = 1/k$, the Hall conductivity law ($J_x = \tfrac{1}{k}E_y$) and the fact ($J_0 = \tfrac{1}{k}(B + j_0)$) that there is one electron per $k$ flux quanta (background and surplus).

Beyond this classical analysis, it is common to assume that, upon quantization, the Chern-Simons term $\tfrac{k}{2} a\wedge \mathrm{d}a$ in \cref{EffectiveCSLagrangian} induces quantum $\mathrm{U}(1)$-Chern-Simons theory (cf. \cite{Manoliu1998}) for the ``statistical gauge field'' $a$ and with it the signatures of abelian topological order (cf. \cite[\S 5.2.3]{Tong2016}).

Even setting aside the issue of artificially dropping the remaining summands in \cref{EffectiveCSLagrangian}, we highlight a conceptual tension in this Ansatz: Since $F$ and $j$ are necessarily integral forms (whose periods count background and surplus magnetic flux quanta, respectively, by Dirac flux quantization of the magnetic field, cf. \cite[\S 2.1]{SS25-Flux}), equation \cref{EffectiveEoM} implies that $J = \mathrm{d}a$ is fractional. While this accommodates the fractional quasi-particle charge that motivates \cref{EffectiveCSLagrangian}, it generally \emph{invalidates} $\tfrac{k}{2}a \wedge \mathrm{d}a$ as a quantum-consistent Chern-Simons term. An aspect of just this problem has been noticed in \cite[p 35]{Witten2016}\cite[p 159]{Tong2016}, whose resolution is admitted to be an ``open question'' in \cite[p 40]{Willsher2020}.

We point out that the problem goes deeper still: Even with the source terms in \cref{EffectiveCSLagrangian} disregarded and with $\mathrm{d}a$ assumed integral and hence quantum-consistent, a more careful analysis of the effective FQH field theory shows that the statistical gauge field generally has a kinetic Maxwell term besides the Chern-Simons term (cf. \cite[(11.41)]{Fradkin2013}\cite[(402)]{Fradkin2024}):
\begin{equation}
  \label{MCSLagrangian}
  L_{\mathrm{MCS}_3}
  :=
  \tfrac{1}{T}
  \mathrm{d}a 
    \wedge 
    \star_3
  \mathrm{d}a
  +
  \tfrac{k}{2}
  a \wedge \mathrm{d}a
  \mathrlap{\,.}
\end{equation}
The Maxwell term is commonly alleged to be ``irrelevant'', and it certainly vanishes in the strong coupling limit $\inlinetikzcd{T \ar[r] \& \infty}$ where the Chern-Simons interaction energy dominates the kinetic energy. 

However, for all finite $T$ the \emph{phase space} of the MCS theory (whose canonical coordinate is $a$) is different from that of CS theory (whose canonical coordinate is only $a_1$ for $\mathrm{d}a = 0$), whence the $\inlinetikzcd{T \ar[r] \& \infty}$ limit of the MCS quantum theory is \emph{different} from plain Chern-Simons theory: it still contains excitation modes, cf. \cite[\S IX]{Haller1996} and \cite{SS26-SDiff}, the latter based on \cite{Pickrell2000}. Beyond these references, this subtle strong coupling limit of MCS theory may not have found due attention.

\subsection{Proper Topological Limit}

This last problem, however, suggests a natural resolution of the issue, as follows: 

In order to avoid taking a problematic limit in a putative space of quantum field theories and instead to take an ordinary limit of observables in a single QFT, we may consider higher-dimensional \emph{5d Maxwell-Chern-Simons} theory, with Lagrangian density (cf. \cite[\S II]{Hill2006})
\begin{equation}
  \label{5DMCS}
  L_{\mathrm{MCS}_5}
  =
  \tfrac{1}{T'}
  \mathrm{d}a'
  \wedge
  \star_5
  \mathrm{d}a'
  + 
  \tfrac{k'}{6}
  a'\wedge \mathrm{d}a'
  \mathrlap{\,,}
\end{equation}
and dimensionally reduce it (cf. \cite[\S 3]{SS25-WilsonLoops}) on a 2-dimensional fiber of volume $V$ carrying flux $\Phi$ (a ``flux compactification''). A standard computation readily shows that the reduced Lagrangian density in 3D is again of the ordinary $\mathrm{MCS}_3$ form \cref{MCSLagrangian}, but now with \emph{dynamically} rescaled parameters:
\begin{equation}
  T = T'/V
  \;\;\;
  \text{and}
  \;\;\;
  k = k' \Phi
  \mathrlap{\,.}
\end{equation}
This way the strong coupling limit in 3D is now recast equivalently as the dimensional reduction limit $\inlinetikzcd{V \ar[r] \& 0}$ in 5D.

\subsection{Proper Flux Quantization}
\label
{OnProperFluxQuantization}

This reformulation of the problem has the profound consequence that we may now securely access the effective formulation of the topological FQH quantum phase by understanding the topological quantum observables of 5D MCS theory.  This has recently been discussed in \cite{SS25-WilsonLoops} (review in \cite{SS25-Complete}, based on \cite{SS25-Flux}).

The key point is that the Gauss laws of 5D MCS  theory are non-linear
\begin{equation}
  \label{GaussLawsOf5DMCS}
  \mathrm{d} b_2 = 0
  \,\;\;\;\;
  \mathrm{d} e_3 = \tfrac{1}{2} b_2 \wedge b_2
  \,,
\end{equation}
which implies \cite{SS24-Phase} that magnetic ($b_2$) and electric ($e_3$) flux densities can no longer be quantized in ordinary 2-cohomology, since that fails to implement the nonlinear relation in \cref{GaussLawsOf5DMCS}. In contrast, the basic admissible flux quantization law in this case is (by \cite[p 40]{SS25-Flux}) \emph{2-Cohomotopy} (cf. \cite[\S VII]{STHu59}\cite[Ex. 2.7]{FSS23-Char}):

This means that where ordinary electromagnetic gauge fields $A$ are classified by maps to the usual classifying space $B \mathrm{U}(1) \simeq \mathbb{C}P^\infty$, the 5D gauge field may be classified by maps to its 2-sphere subspace 
\begin{equation}
  \label{TheClassifyingSpace}
  \substack{
    \text{\scriptsize\color{black!65}classifying space for}
    \\
    \text{\scriptsize\color{black!65}5D MCS fluxes}
  }
  \;
  \mathbb{C}P^1 
    \subset 
  \mathbb{C}P^\infty
  \;
  \substack{
    \text{\scriptsize\color{black!65}classifying space for}
    \\
    \text{\scriptsize\color{black!65}Maxwell fluxes.}
  }
\end{equation}
(The reason for this is that the \emph{minimal Sullivan model} of the space $\mathbb{C}P^1$ reflects exactly the differential relations \cref{GaussLawsOf5DMCS}, cf. review in \cite[\S 3.2]{SS25-Flux} based on \cite{FSS23-Char}.)

Concretely, to model localized (solitonic) flux quanta one needs to impose that flux \emph{vanishes at infinity} (cf. \cite[(2.31)]{Chapman2000}\cite[(6.101)]{Timm2026}), whence we are to consider (cf. \cite[\S 2.2]{SS25-Flux}\cite[\S A.2]{SS24-Obs}):
\begin{enumerate}
\item
the spacetime domain $X_{\mathrm{dom}}$ as pointed by the \emph{point at infinity}, $\infty \in X_{\mathrm{dom}}$, 
\item the charge classifying space $\mathcal{A}$ as pointed by the vanishing charge classifier, $0 \in \mathcal{A}$,
\end{enumerate}
so that the basepoint-preserving classifying maps
\begin{equation}
  \label{BasepointPreservinClassifyingMaps}
  \rchi 
  \in
  \mathrm{Maps}^\ast\bracket({
    X_{\mathrm{dom}}, \mathcal{A}
  })
\end{equation}
literally take the value $0$ at $\infty$:
\begin{equation}
  \label{MapsVanishingAtInfinity}
  \begin{tikzcd}[row sep=-2pt, column sep=0pt]
    X_{\mathrm{dom}}
    \ar[
      rr,
      "{ \rchi }"
    ]
    &&
    \mathcal{A}
    \\
    \infty 
      &\mapsto& 
    0
    \mathrlap{\,.}
  \end{tikzcd}
\end{equation}

For instance, for an FQH liquid occupying a surface $\Sigma^2$, the spatial domain of the uplift to $\mathrm{MCS}_5$ on an $\epsilon$-small fifth dimension interval is pointed homotopy equivalent to
\begin{equation}
  \label{FQHSpacetimeDomain}
  \begin{aligned}
  X_{\mathrm{dom}} 
    & \defneq 
  \bracket({
  \mathbb{R}^1
    \times 
  \Sigma^2 
    \times 
  [-\epsilon, +\epsilon]
  })_{\cpt}
  \\
  & \sim
  S^1 \wedge \Sigma^2_{\cpt}
  \,,
  \end{aligned}
\end{equation}
where ``$\wedge$'' denotes the \emph{smash product} of pointed spaces (their product space with all points identified that are at infinity in one or the other factor).

Hence with this spacetime domain \cref{FQHSpacetimeDomain} and with $\mathcal{A} \defneq \mathbb{C}P^1$ \cref{TheClassifyingSpace}, the topological quantum observables, being the compactly supported functions $\mathbb{C}[-]$ on the set of solitonic field sectors, form the \emph{group algebra}
\begin{equation}
  \label{TheTopologicalSectors}
  \begin{aligned}
  \mathrm{Obs}
  & :=
  \mathbb{C}
  \bracket[{
    \pi_0\, 
    \mathrm{Maps}^\ast\bracket({
      X_{\mathrm{dom}},
      \mathbb{C}P^1
    })
  }]
  \\
  & \simeq
  \mathbb{C}
  \bracket[{
    \pi_1\, 
    \mathrm{Maps}^\ast\bracket({
      \Sigma^2_{\cpt}
      \,,
      \mathbb{C}P^1
    })  
  }]
  \mathrlap{\,,}
  \end{aligned}
\end{equation}
and hence topological quantum states (constituting modules over this algebra) fall into the corresponding irreps (cf. \cite[\S 3]{SS24-Obs}\cite[\S 2]{SS25-Complete}).

\subsection{%
\texorpdfstring
{Relation to Hopfion/$\mathbb{C}P^1$-Model}
{Relation to Hopfion/CP1-Model}
}

In the default case of an FQH liquid on the plane/disk, where $\Sigma^2 \simeq \mathbb{R}^2$ in \cref{FQHSpacetimeDomain} with (by \emph{stereographic projection})
\begin{equation}
  \label{SphereAsCompactificationOfPlane}
  \mathbb{R}^2_{\cpt} \simeq S^2
  \mathrlap{\,,}
\end{equation}
the  topological soliton sectors \cref{TheTopologicalSectors} are labeled by integer multiples of the \emph{Hopf fibration},
\begin{equation}
  \label{HopfClassification}
  \pi_1\,
  \mathrm{Maps}^\ast\bracket({
    \mathbb{R}^2_{\cpt},
    \mathbb{C}P^1
  })
  \simeq
  \pi^3\bracket({S^2})
  \simeq
  \mathbb{Z}
  \mathrlap{\,,}
\end{equation}
as such known from the \emph{Hopfion model} for abelian anyons \cite{Wilczek1983}\cite[\S II.C]{Forte1992}. 

Indeed, closer analysis via the classical \emph{Pontrjagin theorem} (cf. \cite[\S 3.2]{SS23-Mf}) shows \cite{SS25-AbelianAnyons} that the left hand side of \cref{HopfClassification} classifies framed oriented links $L$ (of anyon worldlines) by their \emph{total crossing number} (or \emph{writhe}) $\#L$. This is exactly \cite[Thm. 1]{SS25-WilsonLoops} the exponent of the renormalized expectation value $\langle \nu \vert - \vert \nu \rangle$ of Wilson loop observables in $\mathrm{U}(1)$-Chern-Simons theory at level $k = 1/\nu$:
\begin{equation}
  \begin{tikzcd}[row sep=-2pt, column sep=5pt]
    \mathrm{Map}^\ast\bracket({
      X^3_{\mathrm{dom}}
      ,
      \mathbb{C}P^1
    })
    \ar[
      rr,
      ->>,
      "{
        \pi_0
      }"{yshift=1pt}
    ]
    \ar[
      rrrr,
      uphordown,
      "{
        \langle \nu\vert-\vert\nu\rangle
      }"{description}      
    ]
    &&
    \pi^3\bracket({S^2})
    \ar[
      rr
    ]
    &&
    \mathbb{C}
    \\
    L 
      &\mapsto&
    \#L
     &\mapsto&
    e^{
      \pi \mathrm{i}\nu\,\# L
    }
    \mathrlap{\,,}
  \end{tikzcd}
\end{equation}
reflecting exactly the anyonic braiding phases \cref{TheBraidingPhase}.

\subsection{Covariantized Observables}

However, the topological observables of $\mathrm{MCS}_5$ flux-quantization in 2-Cohomotopy go beyond the Hopfion model. 
First, we may consider other FQH geometries. Notably for the torus, $\Sigma^2 \defneq T^2$, the topological observables \cref{TheTopologicalSectors} turn out to be \cite[Prop. 3.19]{SS25-FQH}\cite[Thm. 3.3]{Kallel2026}:
\begin{equation}
  \label{AlgebraOfObservablesOverTorus}
  \begin{aligned}
  \mathrm{Obs}
  & =
  \mathbb{C}
  \bracket[{
    \pi_1
    \mathrm{Map}\bracket({
      T^2,S^2
    })
  }]
  \\
  &\simeq
  \mathbb{C}\bracket[{
    W_a, W_b,
    \zeta
  }]
  \big/
  \bracket({
    \substack{
    W_a W_b 
    =
    \zeta^{2}
    W_b W_a,
    \\
    \zeta \, \text{central}
    }
  })
  \end{aligned}
\end{equation}
which is exactly the algebra of quantum observables expected for abelian anyons on the torus (\cite[(4.9)]{WenNiu1990}, cf. \cite[(5.28)]{Tong2016}).
That this follows as the fundamental group algebra of maps from the torus to $\mathbb{C}P^1$ is a remarkable fact with deep relevance \cite{SS25-FQAH,SS25-Crys} also for the putative crystalline (``anomalous'') form of FQH anyons. 

But in more detail, topological quantum systems ought to be ``generally covariant'', meaning that any pair of configurations differing only by a diffeomorphism of the spacetime domain ought to be regarded as physically equivalent. For the classifying maps \cref{BasepointPreservinClassifyingMaps}, this means \cite[Def. 2.11]{SS25-FQH} that their physically relevant space is the \emph{homotopy quotient} $(-)\sslash (-)$ (Borel construction) of the mapping space by the action of the diffeomorphism group at least of the surface $\Sigma^2$ \cref{FQHSpacetimeDomain}. Hence, the algebra of \emph{covariantized} topological quantum observables is:
\begin{equation}
  \label{CovariantizedFundamentalGroup}
  \begin{aligned}
  \mathrm{CovObs}
  &
  :=
  \mathbb{C}\bracket[{
  \pi_1
  \bracket({
    \mathrm{Maps}^\ast\bracket({
      \Sigma^2
      ,\,
      \mathbb{C}P^1
    })
    \sslash
    \mathrm{Diff}^+\bracket({
      \Sigma^2
    })
  })
  }]
  \\
  & 
  \simeq
  \mathbb{C}\bracket[{
    \pi_1
    \bracket({
    \mathrm{Maps}^\ast\bracket({
      \Sigma^2
      ,\,
      \mathbb{C}P^1
    })  
    })
    \rtimes
    \mathrm{MCG}\bracket({
      \Sigma^2
    })
    }]
    \mathrlap{\,.}
  \end{aligned}
\end{equation}
As shown in the second line, this entails \cite[Prop. 2.24]{SS25-FQH} passing to the semidirect product of the fundamental group of maps by the surface's \emph{mapping class group} $\mathrm{MCG}\bracket({\Sigma^2}) := \pi_0 \mathrm{Diff}^+\bracket({\Sigma^2})$ (cf. \cite{FarbMargalit2012}).

Remarkably, over the torus $\Sigma^2 \defneq T^2$, the condition that topological quantum states, being finite-dimensional irreps of \cref{AlgebraOfObservablesOverTorus}, are generally covariant in that they extend to representations of \cref{CovariantizedFundamentalGroup}, implies \cite[\S 3.4]{SS25-FQH} the restriction of the braiding phase $\zeta$ \cref{TheBraidingPhase} to a root of unity and the expected ground state degeneracy reflecting the topological order of the quantum system.

This shows that the covariantization of the topological observables \cref{CovariantizedFundamentalGroup} captures crucial physical properties of the FQH liquid. 

\section{Results}

The above discussion (\cref{Methods}) shows that the assumption that surplus FQH flux is quantized in 2-Cohomotopy implies right away the fine detail of the topological order of the corresponding quasi-hole anyons, whose traditional derivation via effective Chern-Simons Lagrangians is less robust (\cref{CritiqueOfEffectiveLagrangians}) and subject to \emph{ad hoc} renormalization choices \cite{SS25-WilsonLoops}.

It is thus compelling to use this novel theory to derive predictions about FQH systems not obviously formulatable with Lagrangian effective field theory.

\subsection{Modelling Flux-Expulsion Islands}

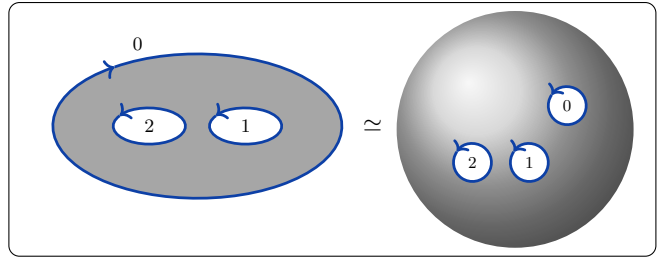
\begin{figure}[htb]
\caption{\label{PuncturedDiskAndSphere}%
  A disk (hosting an FQH fluid) with $n$ islands (punctures) is homeomorphic to a sphere with $n+1$ punctures. From the view of the complement bulk, all punctures are \emph{at infinity} \cref{CompactificationOfPuncturedDisk}, which entails with \cref{MapsVanishingAtInfinity} that magnetic flux does not enter here. 
}
\centering

\adjustbox{
  rndfbox=4pt
}{
\adjustbox{
  raise=-1.3cm,
  scale=.8
}{
    \begin{tikzpicture}
      \draw[
        line width=1.3,
        darkblue,
        fill=gray!70,
        ->
      ]
        (125:2*1.2 and 1*1.2)
        arc
        (125:125-360:2*1.2 and 1*1.2);
     \node at (110:2*1.45 and 1*1.45) 
       {
         $0$
       };

 \begin{scope}[
   shift={(-.8,0)}
 ]
      \draw[
        line width=1.3,
        darkblue,
        fill=white,
        ->
      ]
        (145:2*.3 and 1*.3)
        arc
        (145:145+360:2*.3 and 1*.3);
     \node at (0,0) {${2}$};
  \end{scope}

 \begin{scope}[
   shift={(+.8,0)}
 ]
      \draw[
        line width=1.3,
        darkblue,
        fill=white,
        ->
      ]
        (145:2*.3 and 1*.3)
        arc
        (145:145+360:2*.3 and 1*.3);
     \node at (0,0) {${1}$};
  \end{scope}
\end{tikzpicture}
}
\hspace{-.3cm}
$\simeq$
\hspace{-.1cm}
\adjustbox{
  scale=.7,
  raise=-1.55cm
}{
\begin{tikzpicture}[scale=0.9]

  \shade[ball color=gray!40]
    (0,0) circle (2.5);

\begin{scope}[
    shift={(+.3,-.7)}
  ]
  \draw[
    line width=1.5pt,
    draw=darkblue,
    fill=white,
    ->
  ] 
    (145:.4) arc 
    (145:145+360:.4);
  \node at (0,0) 
  { ${1}$ };
\end{scope}

 \begin{scope}[
    shift={(+.3-1.2,-.7)}
  ]
  \draw[
    line width=1.5pt,
    draw=darkblue,
    fill=white,
    ->
  ] 
    (145:.4) arc 
    (145:145+360:.4);
  \node at (0,0) 
  { ${2}$ };
\end{scope}
 
  \begin{scope}[
    shift={(+1.1,+.5)}
  ]
  \draw[
    line width=1.5pt,
    draw=darkblue,
    fill=white,
    ->
  ] 
    (145:.4) arc 
    (145:145+360:.4);
  \node at (0,0) { $0$ };
  \end{scope}
  
\end{tikzpicture}
}
}
\end{figure}

Consider now the case that $\Sigma^2$ in \cref{FQHSpacetimeDomain} is a \emph{punctured} surface. The experimentally most relevant case is that of the $n$-punctured plane/disk 
\begin{equation}
  \label{PuncturedDisk}
  \Sigma^2_{0,n}
  :=
  \mathbb{R}^2
  \setminus 
  \{ x_1, \cdots, x_n \}
  \,.
\end{equation}
Mathematically, this is a standard choice of surface to plug into the general formula for topological FQH observables found in \cref{CovariantizedFundamentalGroup}. Let us see what this mathematical model describes physically:

The key observation here is that each puncture in \cref{PuncturedDisk} is an ``asymptotic boundary'' (being the complement of a closed point), which implies that each of the points deleted in \cref{PuncturedDisk} is ``at infinity'' as seen from $\Sigma^2_{0,n}$. This entails that adjoining infinity as a formal new point to the spacetime domain \cref{FQHSpacetimeDomain} amounts to identifying all these points with the single point-at-infinity, hence is the result of \emph{pinching} the sphere \cref{SphereAsCompactificationOfPlane} at $n+1$ points (cf. \cref{PuncturedDiskAndSphere}):
\begin{equation}
  \label{CompactificationOfPuncturedDisk}
  \bracket({
    \Sigma^2_{0,n}
  })_{\cpt}
  \simeq
  S^2 / 
  \{
    x_0, \cdots, x_n
  \}
  \,.
\end{equation}

To see the physical meaning of this mathematical result, recall that the flux classifying maps \emph{vanish at infinity} \cref{MapsVanishingAtInfinity}. Before we punctured $\mathbb{R}^2$, this encoded just the evident physical condition that the magnetic field is threaded through the FQH liquid under consideration, not extending beyond its boundary. Including these punctures models a scenario in which \emph{magnetic flux is expelled also from the vicinity of each of the punctures}.

Note that for instance a gate-depleted antidot or similar common defect may remove the liquid but not the flux, hence need not realize \cref{MapsVanishingAtInfinity}. Instead, the canonical laboratory realization of the required (surplus) flux-expulsion is a superconducting island in the FQH liquid (cf. \cite{Lindner2012,clarke2013,Mong2014,GulEtAl2022,CaoEtAl2026}), via its Meissner effect (which will have to be stabilized against the strong transverse magnetic field). We stress, however, that the following depends only on the flux expulsion as such and not on the microscopic physics realizing it, so that superconducting islands are only one possibility. 

Therefore, using the $n$-punctured disk \cref{PuncturedDisk} in the general formula \cref{CovariantizedFundamentalGroup}  yields the topological observables, and hence the topological quantum states, of planar FQH liquids hosting $n$ flux-expulsion islands.

\subsection{Deriving the Anyon Attachment}

\begin{figure}[htb]
  \caption{\label{ASphericalBraid}%
    \emph{Spherical braids} \cref{TheSphericalBraidGroup} are braids of worldlines of punctures on the 2-sphere (cf. \cref{PuncturedDiskAndSphere}). These are almost the same as ordinary Artin braids of punctures in the plane. (Shown is the element $b_1 b_2 b_1 b_2 b_1$). The only difference being that on the sphere, the Artin braid $(b_1 \cdots b_n)(b_n \cdots b_1)$  (where all punctures are fixed except one, which is circling the others) is topologically trivial.  
  }
  \centering
  \adjustbox{
    rndfbox=4pt
  }{
    \hspace{3pt}\includegraphics[width=.65\linewidth]{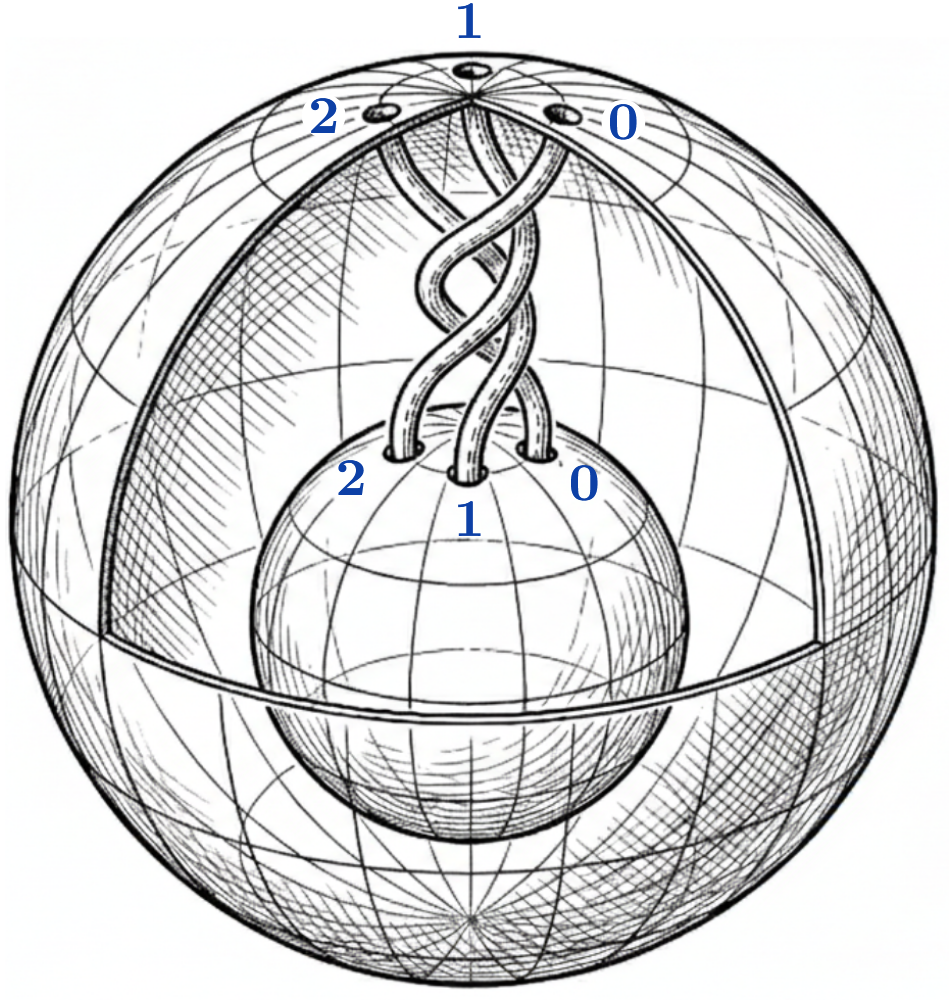}
  }
\end{figure}

We have thereby reduced the physical analysis to the question of computing the topological observable algebra obtained from plugging \cref{PuncturedDisk,CompactificationOfPuncturedDisk} into the general formula \cref{CovariantizedFundamentalGroup} 
\begin{equation}
  \label{ObservablesOnPuncturedDisk}
  \begin{aligned}
  &\mathrm{CovObs}
  \simeq
  \\
  &
  \mathbb{C}\bracket[{
    \pi_1
    \mathrm{Maps}^\ast\scaledbracket({
      \bracket({
        \Sigma^2_{0,n}
      })_{\cpt}
      ,
      \mathbb{C}P^1
    })
    \rtimes
    \mathrm{MCG}\scaledbracket({\Sigma^2_{0,n}})
  }]
  \mathrlap{\,.}
  \end{aligned}
\end{equation}
Computing this is now a problem purely in low-dimensional algebraic topology and may be attacked with classical tools of that field. The computation is spelled out in \cite[\S 3.5-6]{SS25-FQH} with the result that:
\begin{proposition}
  \label[proposition]{AppearanceOfBraidGroup}
  The topological quantum states on the $n$-punctured disk according to \cref{ObservablesOnPuncturedDisk} fall into unitary irreps of the subgroup of the $\mathrm{rot}$-quotient of the \emph{framed spherical braid group} with $n+1$ strands
  \begin{equation}
    \label{TheFramedBraidGroup}
    \mathrm{FBr}_{n+1}(S^2)/\mathrm{rot}
    :=
    \mathbb{Z}^{n+1}
    \rtimes 
    \mathrm{Br}_{n+1}(S^2)/\mathrm{rot}
  \end{equation}
  on framed braids whose total framing number is a multiple of $n+1$.
\end{proposition}
Here the formula \cref{TheFramedBraidGroup} unwinds as follows:
\begin{enumerate}
\item
 One factor of $\mathbb{Z}$ in $\mathbb{Z}^{n+1}$ arises as the Hopfion contribution in  \cref{HopfClassification}, accompanied by further factors for each island, signifying that solitonic abelian anyons now have one separate braiding phase \cref{TheBraidingPhase} for braiding in the vicinity of each island. 

\item But the striking novel import of \cref{TheFramedBraidGroup} is the appearance of the \emph{nonabelian} (for $n > 1$) spherical braid group $\mathrm{Br}_{n+1}(S^2)$ (cf. \cite[Lit. 2.20]{MySS2024}\cite[Ex. 2.17]{SS25-FQH}), formally reflecting the motion of possibly nonabelian defect anyons around each other (cf. \cite[\S II.A]{Nayak2008}), here associated with the (surplus) flux-expulsion islands. 

\item Specifically, the spherical braid group appearing here (cf. \cref{ASphericalBraid}) is the quotient of the usual Artin braid group $\mathrm{Br}_{n+1}$ (with its set of Artin generators $\{b_i\}_{i=1}^n$) by the braid where one strand goes once around all others (cf. \cite[p. 245]{FadellVanBuskirk1961}):
\begin{equation} 
  \label{TheSphericalBraidGroup}
  \mathrm{Br}_{n+1}(S^2)
  \simeq
  \mathrm{Br}_{n+1}
  \big/
  \bracket({
   \bracket({
     b_1 \cdots b_n
   }) 
   \bracket({
     b_n \cdots b_1
   })
  })
  \mathrlap{\,.}
\end{equation}

\item Furthermore, the element $\mathrm{rot}$ (cf. \cite[(Fig. 9.6)]{FarbMargalit2012}) denotes the braid whose strands jointly perform a full rotation. Its square turns out to be equivalent to the trivial braid and hence the expression $(-)/\mathrm{rot}$ in \cref{TheFramedBraidGroup} means the quotient by the corresponding $\mathbb{Z}_{/2}$ subgroup.

\item These two quotients mean that the nonabelian braiding operations appearing here are a little more constrained than usually assumed for nonabelian anyons. For example, for $n = 2$ the monodromy \cref{TheFramedBraidGroup} reduces to the \emph{framed symmetric group} (cf. \cite[(42)]{SS25-FQH})
\begin{equation}
  \mathrm{FBr}(S^2)/\mathrm{rot}
    \simeq
  \mathbb{Z}^3 
    \rtimes
  \mathrm{Sym}_3
\end{equation}
and hence exhibits the special case of ``parastatistical'' anyons (cf. \cite[Rem. 3.60]{SS25-FQH}). These may deserve more attention as they still implement non-Clifford quantum gates (cf. \cite[Prop. 3.61]{SS25-FQH}).

\end{enumerate}

While \cref{AppearanceOfBraidGroup} does not specify the spherical braid representation in which the ground states will transform (which will depend on microscopic details beyond the presence of flux-expulsion islands), it shows that there may be superselection sectors (irreps) beyond the abelian phase which exhibit nonabelian anyonic braid statistics, in contrast to the abelian situation \cref{HopfClassification} without flux-expulsion islands.

\section{Conclusion}

Given that abelian anyons in 2D FQH liquids have been experimentally observed by independent groups, while the popular idea of engineering nonabelian anyons on super/semi-conductor heterostructures remains experimentally inconclusive in 1D,  
here we revisited the theory of flux-expulsion islands in 2D FQH liquids, possibly via semi/superconducting heterostructures as previously explored by only a few authors.

Observing that these strongly correlated electron systems are non-perturbative and not as reliably captured (effectively at long wavelengths) by 3D Chern-Simons Lagrangian densities as often assumed, our key move was to invoke instead a recently proposed alternative effective description \cite{SS25-WilsonLoops,SS25-AbelianAnyons,SS25-FQH}: 

Here the ground-state monodromy characterizing anyonic topological order is directly modeled by representations of fundamental groups of topological parameter spaces of maps from the material's sample to a deformation of the usual electromagnetic classifying space $\mathbb{C}P^\infty$. 

Taking the deformed classifying space to be the 2-skeleton $\mathbb{C}P^1 \subset \mathbb{C}P^\infty$, thought of as the effective quantization law for the surplus FQH flux quanta (vortex quasi-holes), turns out to accurately and reliably retrodict fine detail of the familiar abelian FQH anyon states, thus lending weight to its implications for the effect of flux-expulsion islands in FQH liquids.

To this end, using computations in low-dimensional algebraic topology which extend the old Hopfion model for abelian anyons, 
we arrive at the striking result (\cref{AppearanceOfBraidGroup}) that, besides a proliferation of abelian braiding phases, the system's ground states now fall into irreps of (a mild quotient of) the framed spherical braid group. Since this group is nonabelian (for $n > 1$), such irreps may be higher-dimensional, in which case the islands carry nonabelian anyons. Of course, which irrep is actually realized depends on microscopic details of the island and its boundary and is left to future work.

While it may be challenging to engineer flux-expulsion islands in FQH liquids, even more so with controlled mobility, our result suggests that this scenario may be a worthwhile target for experiment to explore.


\medskip

\noindent
{\bf Declarations.}
The authors declare that they have no conflict of interest. No data were generated during this theoretical research.

\noindent
{\bf Funding statement.}
This research was supported by \emph{Tamkeen UAE} under the 
\emph{NYU Abu Dhabi Research Institute grant} \texttt{CG008}.

\bibliography{refs.bib}

\end{document}